\documentclass{article}
\usepackage[paperwidth=612pt,paperheight=792pt,top=72pt,right=72pt,bottom=72pt,left=72pt]{geometry}
\usepackage{graphicx}
\usepackage{dcolumn}
\usepackage{bm}
\usepackage{hyperref}
\usepackage{amssymb}
\usepackage{amsmath}
\usepackage{latexsym}
\begin{document}
\title {Quark confinement due to creation of micro AdS black holes in quarkonium model}
\author{ M. Taki $^{a}$\footnote{metaki@stu.yazd.ac.ir},A. Mirjalili $^{a}$\footnote{a.mirjalili@yazd.ac.ir} }
\maketitle
{\begin{center}{
$^{a}$ Physics Department, Yazd University, 89195-741,Yazd, Iran}\end{center}}

\begin{abstract}We use the solution of  the Dirac equation for quarkonium atom in the 4D Anti de sitter (AdS$_{4}$) space to investigate the effect of the large negative cosmological constant
on the phenomenon of quark confinement. We do the required calculations in the  AdS$_{4}$ space to indicate that   large cosmological constant can describe the quark confinement. In fact using the coulomb potential in Dirac equation while we employ the AdS metric will additionally lead us   to a linear potential  in the quark-antiquark interaction which can be considered to explain the quark confinement.  This
confining term is arising  essentially  from the geometrical features of the space.  On the other hand the origin of the large
cosmological constant can be justified by assuming  the appearance of micro black holes in the recent  hadronic collision process which is now current, for instance, at the LHC project.
\end{abstract}


\section{Introduction}
It has  been known for long time that due to quark confinement and the short range of the strong force, the production of quark anti-quark pair in the $e^+$ $e^-$ collision will result to hadronic jets instead of individual colored quarks. But the mechanism of this confinement is still ambiguous. There were many attempts to explain the colour confinement such as bag models \cite{{Q1,Q2,Q3,Q4}}, confinement potential models \cite{Q5,Q6}, quasi particle models
\cite{Q7}, strongly interacting quark-gluon-plasma (sQGP) \cite{{Q8,Q9,Q10}} and etc. In spite the many attempts which have been done but  people still do not reach to an unique and acceptable  framework on this issue.\\

In this work we investigate the effect of the four dimensional (4D) Ads metric on the quarkonium atom that its coloured quark-antiquark interaction is described only by Coulomb-like potential $V\left( r \right)=\frac{a}{r}$. As we will see the existence of an Ads metric with a negative cosmological constant
impose the confining term $\frac{\Lambda }{6}ar$ in the Dirac equation which is added to the coulomb potential term. In comparison with other confining
models, the new effective potential $V_{eff} \left( r \right)=\frac{a}{r}-\frac{\Lambda }{6}ar$ acts like the Cornell potential
\cite{Q11}. Cornell potential models that are used to describe the mass spectrum  of heavy mesons have been established on a non-relativistic potential
$V_{Cornell} \left( r \right)=\frac{a}{r}-br$ , where its Coulomb part is a non-relativistic limit of the one-gluon exchange interaction dominant at
short distance interactions, while the linear part comes from the Wilson loop and ensures the confinement of quarks. As an interesting result,
using the Ads metric we achieve  the quark confinement  in a hydrogen-like atom which is actually resulted from the geometrical properties of space.\\

By comparing the unknown constants  in effective potential with the Cornell potential whose parameters are determined via  fitting  the experimental data, we obtain the numerical
value of cosmological constant as $\frac{\Lambda }{6}\approx 10^{30}m^{-2}$. Such a large cosmological constant can be  produced by a black holes in
its nearby space. By new hypotheses,  micro black holes could be performed at an energy scale of the $TeV$ order \cite{Q12} which are available now in recent colliding machines at CERN. When one try to de-confine the quarks in the quarkonium atom then the creation of Ads micro black holes is expected \cite{Q13,Q14}.\\

It is known that for a Schwarzschild black hole in an AdS space, the black hole is thermodynamically unstable when the horizon radius is small \cite{Q16}.
So, in contrast with normal black holes that have a long life time, micro black holes have a very short life time of $10^{-26}s$ order. This means that
once they will be created they will evaporate instantaneously. However, as we will discuss the creation of AdS micro black holes while trying to de-confine the quarks can
explain the quark confinement.\\

On the other hand, there is an inspiring and different idea which was proposed by E.Witten that tried to explain the connection between geometry of space and the quark
confinement problem via the Ads space \cite{Q15}. This  can be considered as a confirmation to what we do to employ the Ads metric to extract the linear potential which is justifying the quark confinement. At the present time, the developed AdS/QCD approach to hadron physics provides  remarkable
descriptions in many phenomenological features of QCD.\\

The outline of the paper is as follows. In section 2, we introduce the Dirac Equation on Anti de sitter space. The Solution of Dirac equation
for  quarkonium atoms, using the Ads metric is described in Section 3 which lead us to an additional  linear potential. We give finally our conclusion in section 4.
\section{The Dirac equation on AdS metric}

The AdS$_{4}$ metric in a global coordinates is defined by
\begin{equation}
\label{eq1}
g_{\mu \nu } dx^\mu dx^\nu =F\left( r \right)dt^2-\frac{1}{F\left( r
\right)}dr^2-r^2\left( {d\theta ^2+\sin ^2\theta \mbox{ }d\phi ^2} \right),
\end{equation}
where $F\left( r \right)=1+\frac{\Lambda }{3}r^2$ \cite{Q16}.

We consider the covariant form of Dirac equation on the metric $g_{\mu
\nu } $ that has the form
\begin{equation}
\label{eq2}
i\gamma _{\left( g \right)}^\mu \nabla _\mu \psi -m\psi =0\;.
\end{equation}
In this case $\gamma _{\left( g \right)}^\mu $ are the contravariant Dirac matrices and $\nabla _\mu $ is the covariant derivative which are satisfying  the following
relations \cite{Q17,Q18}:
\begin{equation}
\label{eq3}
\begin{array}{l}
 \gamma _{\left( g \right)}^\mu =\gamma _\alpha e^{\alpha \mu }\mbox{
}\alpha =1,2,3, \\\\
 \nabla _\mu =\partial _\mu +\frac{1}{8}\left[ {\gamma _\alpha ,\gamma
_\beta } \right]\omega _\mu ^{\alpha \beta }\;, \\
 \end{array}
\end{equation}
where $\gamma _\alpha $ is the ordinary Dirac matrices such that:
\begin{equation}
\label{eq4}
\begin{array}{l}
 \gamma ^0=\left( {{\begin{array}{*{20}c}
 I \hfill & 0 \hfill \\
 0 \hfill & {-I} \hfill \\
\end{array} }} \right),\mbox{ }\gamma ^\alpha =\left(
{{\begin{array}{*{20}c}
 0 \hfill & {\sigma ^\alpha } \hfill \\
 {-\sigma ^\alpha } \hfill & 0 \hfill \\
\end{array} }} \right),
 \end{array}
\end{equation}
in which $\sigma ^\alpha$ and $I$  are 2 $\times$ 2  Pauli and identity matrices respectively. The coefficient of spin connection $\omega _\mu ^{\alpha \beta } $ as a
function of vierbein fields $e_\alpha ^\mu $~is given by \cite{Q17}
\begin{equation}
\label{eq5}
\omega _\mu ^{\alpha \beta } =\frac{1}{2}e^{\alpha \nu }\left( {e_{\nu ,\mu
}^\beta -e_{\mu ,\nu }^\beta } \right)-\frac{1}{2}e^{\beta \nu }\left(
{e_{\nu ,\mu }^\alpha -e_{\mu ,\nu }^\alpha } \right)+\frac{1}{2}e^{\alpha
\nu }e^{\beta \sigma }\left( {e_{\nu ,\sigma }^\rho -e_{\sigma ,\nu }^\rho }
\right)e_{\rho \mu } .
\end{equation}
However, if we choose the local orthonormal Lorentz frame, the Dirac equation with electric potential $V(r)$ has the following form in spherical
coordinate system [see e.g. \cite{Q18}].
\begin{equation}
\label{eq6}
\left[ {iF^{-\frac{1}{2}}\gamma ^0(\frac{\partial }{\partial
t}+iV(r))+iF^{\frac{1}{2}}\gamma ^1\left( {\frac{\partial }{\partial
r}+\frac{1}{r}+\frac{{F}'}{4F}} \right)+\frac{i}{r}\gamma ^2\left(
{\frac{\partial }{\partial \theta }+\frac{1}{2\tan \theta }}
\right)+\frac{i}{r\sin \theta }\gamma ^3\frac{\partial }{\partial \phi }-M}
\right]\psi \left( {r,\theta ,\phi ,t} \right)=0.
\end{equation}
Now, if we put $\psi \left( {r,\theta ,\phi ,t} \right)=\left( {r\left(
{\sin \theta } \right)^{\frac{1}{2}}F^{\frac{1}{4}}} \right)^{-1}\varphi
\left( {r,\theta ,\phi ,t} \right)$, the Eq. (\ref{eq6}) can be rewritten in a simple form as it follows
\begin{equation}
\label{eq7}
\left[ {iF^{-\frac{1}{2}}\gamma ^0(\frac{\partial }{\partial
t}+iV(r))+iF^{\frac{1}{2}}\gamma ^1\left( {\frac{\partial }{\partial r}}
\right)+\frac{i}{r}\gamma ^2\left( {\frac{\partial }{\partial \theta }}
\right)+\frac{i}{r\sin \theta }\gamma ^3\frac{\partial }{\partial \phi }-M}
\right]\varphi \left( {r,\theta ,\phi ,t} \right)=0.
\end{equation}
It is convenient to make a separation of variables in Eq. (\ref{eq7}),  choosing
\begin{equation}
\label{eq8}
\varphi \left( {r,\theta ,\phi ,t} \right)=\frac{1}{r}\left(
{{\begin{array}{*{20}c}
 {f(r)Y_{lm} (\theta ,\phi )} \hfill \\
 {ig(r)Y_{{l}'m} (\theta ,\phi )} \hfill \\
\end{array} }} \right)e^{-iEt}.
\end{equation}
By substituting Eq. (\ref{eq8}) into Eq. (\ref{eq7}) and considering this fact that the angular part of the Dirac equation in AdS space, given by Eq. (\ref{eq1}), is the same as Dirac equation in usual space, we will arrive at
\begin{equation}
\label{eq9}
\begin{array}{l}
 \left[ {M-F^{-\frac{1}{2}}\left( {E-V\left( r \right)} \right)}
\right]f\left( r \right)=F^{\frac{1}{2}}{g}'\left( r \right)-\frac{\kappa
}{r}g\left( r \right), \\\\
 \left[ {M+F^{-\frac{1}{2}}\left( {E-V\left( r \right)} \right)}
\right]g\left( r \right)=F^{\frac{1}{2}}{f}'\left( r \right)+\frac{\kappa
}{r}f\left( r \right), \\
 \end{array}
\end{equation}
where $\kappa \mbox=\pm \left( {j+1/2} \right)$  \cite{Q12}.
\section{The Solution of Dirac equation for  quarkonium atom and quark confinement}

Considering the short distance behavior of quark-antiquark interaction with order of $r\sim 10^{-16}m$ , we then are able to solve the Eq. (\ref{eq9}). For
this purpose, we use the Taylor expansion of  $F^{\frac{1}{2}}$ and $F^{-\frac{1}{2}}$ terms in Eq. (\ref{eq9}) as it follows
\begin{equation}
\label{eq10}
\begin{array}{l}
 F^{\frac{1}{2}}=\left( {1+\frac{\Lambda }{3}r^2}
\right)^{\frac{1}{2}}\approx 1+\frac{\Lambda }{6}r^2-\frac{\Lambda
^2}{72}r^4+{\rm O}(r^5), \\\\
 F^{-\frac{1}{2}}=\left( {1+\frac{\Lambda }{3}r^2}
\right)^{-\frac{1}{2}}\approx 1-\frac{\Lambda }{6}r^2+\frac{\Lambda
^2}{24}r^4+{\rm O}(r^5). \\
 \end{array}
\end{equation}
Substituting these expansions into Eq. (\ref{eq9}) and  ignoring second and higher order terms with respect to $r$ while using Coulombian-like potential $V\left( r
\right)=\frac{a}{r}$ for quark-antiquark interaction, we will get
\begin{equation}
\label{eq11}
\begin{array}{l}
 \left[ {M-\left( {E-\left( {\frac{a}{r}-\frac{\Lambda }{6}ar} \right)}
\right)} \right]f\left( r \right)={g}'\left( r \right)-\frac{\kappa
}{r}g\left( r \right)\;, \\\\
 \left[ {M+\left( {E-\left( {\frac{a}{r}-\frac{\Lambda }{6}ar} \right)}
\right)} \right]g\left( r \right)={f}'\left( r \right)+\frac{\kappa
}{r}f\left( r \right)\;.\mbox{ } \\
 \end{array}
\end{equation}
The appearance of the  $\frac{\Lambda }{6}ar$ term that has been added to the  Coulomb potential in Eq. (\ref{eq11}) is equal to confinement term in Cornell
potential which is used to describe the quark confinement that involves  the Coulomb potential  ``$\frac{a}{r}$" plus a linear term ``$br$"  i.e.
\begin{equation}
\label{eq12}
V_{Cornell} \left( r \right)=\frac{a}{r}-br,
\end{equation}
where $a$ is a parameter representing the Coulomb strength, and $b$ measures the strength of the linear confining term. Now, by comparing the
Cornell potential with the confinement potential that is induced by the AdS metric on Dirac Eq.(\ref{eq11}), one can reach to an expression for cosmological constant $\Lambda $ in terms of potential constants $a$ and $b$ in Eq. (\ref{eq12})
as following
\begin{equation}
\label{eq13}
\frac{\Lambda }{6}=\frac{b}{a}\;.
\end{equation}
Finding the proper value of  $a$ and $b$ parameters in Cornell potential that produce the spectrum of quarkonium atom, had been the main subject of many
works \cite{Q19,Q20,Q21,Q22}. In spite of small differences between the value of parameters $a$ and $b$ in different works, in all of them the ratio $\frac{b}{a}$ is about  $4\times 10^{30}m^{-2}$. So, we can obtain the approximate value of  $\frac{\Lambda }{6}$ term, i.e.
\begin{equation}
\label{eq14}
\left| {\frac{\Lambda }{6}} \right|\approx 4\times 10^{30}m^{-2}.
\end{equation}
In addition, the radial function $f\left( r \right)$ in  Eq.(\ref{eq11}) for Cornell potential can be defined as \cite{Q12}
\begin{equation}
\label{eq15}
f\left( r \right)=Ai\left( {\left( {\frac{\Lambda }{6}a} \right)^{1/3}+a_n }
\right)P_{{n}'} \left( r \right)e^{-\eta r}r^l\;,
\end{equation}
where $Ai(r)$ is the Airy function and  $a_n $ are its roots. In Eq.(\ref{eq15}) $P_{{n}'} \left( r \right)$ is representing the Legendre function and $\eta $ is a function of
quarkonium energy states which is discussed in \cite{Q12}.\\

It is obvious that such a large cosmological constant  can be just produced by a black holes in its nearby space. Furthermore, by new hypotheses that
micro black holes could be performed at the energy scales of  $TeV$ order which are appropriate to destroy the   quarkonium systems, one can expect the creation of
AdS micro black holes to describe the quark confinement through the discussed method.\\

To check the possibility of the creation of micro black holes inside the quarqonium atom we should calculate the horizon radius of black hole that that is imposed by  the  cosmological constant, determined in above.  For the D-dimensional AdS black holes  the maximum event horizon is defined by \cite{Q23,Q24}

\begin{equation}\label{eq16}
r^{2} _{max}=-\left({\frac{D-3}{D-1}}\right) \frac{3}{\Lambda}.
\end{equation}

For $D=4$ and $\frac{\Lambda }{6}\approx 4\times 10 ^{30} m^{-2}$ the maximum value of event horizon radius is about $r _{max}\approx 2 \times 10 ^{-16} m$. By comparing this radius with the radius of the mesons which is detrmined through the de-confining process, we can conclude that  the creation of micro black holes inside mesons is completely possible.

\section{Conclusion }

The main purpose of our work was the investigation of the effect of negative cosmological constant in the Anti de sitter space on the binding states of
quarkonium atoms. We proved that the  4D Ads metric with large cosmological constant can explain the quark confinement by imposing a confining term in
the quark-antiquark interaction potential.\\

Furthermore we discussed that in the case of large cosmological constant, the appearance of micro black holes could be considered as an alternative
explanation for quark confinement phenomenon in the hadron collision. To get the exact solution of the Dirac equation, using the Ads metric for quarkonium atom is a valuable subject which we hope to report on them  as our new research task in future.\\

We should  mention  that in spite to ignore the higher order terms in Eq.(\ref{eq9}),  this
assumption do not disturb our final result. In fact the appearance of extra power terms in the right side of Eq.(\ref{eq9}) can be absorbed into wave function
while these extra terms in the left side of this equation  play the role  of harmonic terms in effective potential. This means that  we  will have a stronger  confining potential with respect to the Cornell potential which are now containing two  linear and quadratic order of $r$ terms  \cite{Q23,Q24}. Investigating this confining potential and the results which are arising it, can be considered as  our further research activity.

\end{document}